\documentclass[twocolumn,showkeys,showpacs,amsmath,amssymb,superscriptaddress,preprintnumbers,nofootinbib]{revtex4}
\usepackage{amssymb}
\usepackage{amsmath}
\usepackage{graphicx}
\usepackage{euscript}

\normalbaselines
\begin{document}

\title{Magneto-electro-statics of axionically active systems: \\ Induced field restructuring in magnetic stars}

\author{Alexander B. Balakin}
\email{Alexander.Balakin@kpfu.ru}
\author{Dmitry E. Groshev}
\email{groshevdmitri@mail.ru}
\affiliation{Department of General Relativity and
Gravitation, Institute of Physics,Kazan Federal University, Kremlevskaya street 18, 420008, Kazan,
Russia}

\newcommand{\nablab}{{\mathop {\rule{0pt}{0pt}{\nabla}}\limits^{\bot}}\rule{0pt}{0pt}}

\begin{abstract}
In the framework of the Einstein-Maxwell-axion theory we consider a static field configuration in the outer zone of a magnetic star. We assume that this field configuration is formed by the following interacting quartet: a spherically symmetric  gravitational field of the Reissner-Nordstr\"om type,  pseudoscalar field, associated with an axionic dark matter, strong intrinsic magnetic field, and axionically induced electric field. Based on the analysis of the solutions to the master equations of the model, we show that the structure of the formed field configuration reveals the following triple effect. First, if the strong magnetic field of the star has the dipole component, the axionic halo around the magnetic star with spherically symmetric gravitational field is no longer spheroidal, and the distortion is induced by the axion-photon coupling. Second, due to the  distortion of the axionic halo the magnetic field is no longer pure dipolar, and the induced quadruple, octupole, etc., components appear thus redistributing the magnetic energy between the components with more steep radial profiles than the dipolar one. Third, the interaction of the axionic and magnetic fields produces the electric field, which  inherits their multipole structure. We attract attention to possible applications of the predicted field restructuring to the theory of axionic rotating magnetars.
\end{abstract}



\maketitle

\section{Introduction}

The term {\it magnetic stars} relates to the wide class of astrophysical objects, rotating, or static, for which the magnetic field plays a fundamental role in their observation and recognition. Among them, the most interesting are the relativistic pulsars and magnetars, which are characterized by extremely high magnetic and gravity fields and which are, in fact, unique cosmic laboratories to probe modified theories of gravity and extended electrodynamics in the strong field limit \cite{ms1, ms11}. Properties of these extraordinary astrophysical sources are the subject of numerous systematic investigations (see, e.g., the reviews \cite{ms2,ms3,ms4} for details and references). By studying the general properties of extended models of magnetic stars we would like to clarify the following question: can predictions of the  axion electrodynamics (axionic extension of the Faraday-Maxwell theory)  be tested using the data extracted from the magnetic star observations?

The key element of the mentioned extended theories is the {\it axion}. It appeared in the theory of fundamental interactions as a new light massive pseudo-Goldstone boson \cite{Weinberg,Wilczek}, due to the theoretical discovery of Peccei and Quinn  of the CP-invariance conservation \cite{PQ}. Axions are considered in the context of explanation of the dark matter phenomenon \cite{ADM1}-\cite{ADM9}, thus providing the interest to the cosmological and astrophysical applications of the axionic extensions of strong field models.  One of the most interesting feature of the axion-photon  coupling is, without a doubts, the following: a magnetic field in a non-uniform axion environment produces an electric field, if the four-gradient of the pseudoscalar field, associated with the axions, is nonvanishing. The term non-uniform includes various states of the pseudoscalar (axion) field $\phi$. For instance, in the cosmological context, the axion field is considered as a function of the cosmological time only, $\phi(t)$, and the corresponding four-gradient $\nabla_i \phi$ is the timelike four-vector, $\nabla_i \phi  \nabla^i \phi {=} {\dot{\phi}}^2>0$.
Respectively, the electric field, induced by constant magnetic field surrounded by relic cosmic dark matter axions, is a function of time and inherits the information concerning the rate
of the Universe expansion (see, e.g., \cite{relic1,relic2,relic3,relic4}). When we deal with the gravitational pp-waves, the pseudoscalar (axion) field can be considered as a function of retarded time $u=(ct{-}x)/\sqrt2$, and the corresponding gradient four-vector is the null one, i.e.,  $\nabla_i \phi  \nabla^i \phi {=} 0$. An electric field, induced due to the axion coupling to the static magnetic field under the influence of the gravitational pp-waves, is shown to reveal an anomalous character (see, e.g., \cite{gw1,gw2,gw3}).

Static field configurations can be characterized by the space-like gradient four-vector with negative square $\nabla_i \phi  \nabla^i \phi < 0$. The structure of the electric field, induced by the magnetic field in the inhomogeneous axionic environment, inherits the complexity of the magnetic field. The simplest static model is the axionic magnetic monopole, in which all fields are assumed to depend on the radial variable only. The corresponding magnetic field has the radial component only, the axionically induced electric field is also radial, and the field configuration relates to the so-called axionic dyon \cite{Wilczek2,LW1991,BaZa2017,BG2019}. The model of the axionic dyon is self-consistent; high order moments of the electric and magnetic fields do not appear. In this sense the model of axionic dyon is unique. When we consider, at least, the dipolar magnetic field, thus introducing not only the radial, but the angular variables also into the model, we obtain automatically that all multipoles become nonvanishing. The most interesting models, of course, are the models with dipole and quadruple moments, since they, in principle, can be tested using spectroscopic techniques. The multipole structure of the electromagnetic field for the case of relic dark matter axions was analyzed in \cite{BaGru} on the example of a weak gravitational field of the Earth. Now we consider the relativistic model of magnetic star with strong magnetic, gravitational, pseudoscalar (axion) fields and axionically induced electric field.

The most known representatives from the class of relativistic magnetic stars are the rotating objects, and their analysis has to be based on the spacetime of the Kerr type. However, many specific features of the axionic extensions of the relativistic magnetic star theory can be predicted based on the static models with the gravitational field of the Reissner-Nordstr\"om type.
Why does the static model seem to be interesting? There are at least three motives to attract attention to this model.

The first motif is connected with the idea of the formation of a distorted axionic dark matter halo in the magnetic star environments. Indeed, we get used to fact that the dark matter halos are spherically symmetric, when the gravity field of the central body is spherically symmetric. The question arises: Is it possible to find a distorted halo, e.g., with dipole and/or quadruple structure, if the halo surrounds the spheroidal body? The answer is yes: below we show that the magnetic stars can form the distorted axionic dark matter halos due to the axion-photon coupling.

The second motif is connected with the extension of ideas of the magnetic star spectroscopy. A strong magnetic field produces the well-known Zeeman effect in atoms in the magnetic star vicinity \cite{Zeeman}; if the axionically induced electric field appears, one can try to find the fingerprints of the Stark effect.  Moreover, one can find the specific zones in the vicinity of the magnetic star, in which the electric and magnetic fields are parallel, and are crossed. The corresponding methods to observe fine effects in the parallel and crossed electric and magnetic fields are well-elaborated in the terrestrial laboratories (see, e.g., \cite{ZeeStark}); clearly, it can be applied to the magnetic star spectroscopy also. In other words, the analysis of the magnetic star spectrum plus detailed prognosis concerning the distribution of the axionically induced electric field, are expected to provide the diagnostics of the axion field structure near the magnetic star.

The third motif is inspired by observations of some enigmatic bursts in such cosmic objects and jump-type variations of their rotation velocities. One of the explanations is connected with a restructuring of the strong magnetic field of the star. It is well known, for instance, that far from the center of the object the quadruple component of the magnetic field potential tends to zero as $r^{-3}$, i.e., more steeply than the dipolar component, characterized by the asymptote $r^{-2}$. In other words, the quadruple magnetic structure is more compact than the dipolar one, the octupole magnetic structure is more compact than the quadruple one, etc. If the relativistic magnetic star rotates rapidly (with the frequency $\Omega$) and the radius of the so-called light-cylinder ${\cal R}_{\rm L}=\frac{c}{\Omega}$ becomes of the same order as the typical size of the magnetic structure ${\cal R}_{\rm M}$, a spontaneous compactification of the magnetic structure is possible, which is accompanied by the energy release and acceleration of rotation. In principle, there are two ways for the magnetic field restructuring: the first way  can be realized as a catastrophic destruction of the monopole and dipole type magnetic field; this way can be connected with the bursts. The second way can look like the series of mild energy redistributions between, e.g., the dipole and quadruple, quadruple and octupole moments, etc.
In order to recognize the way of restructuring, we have to know whether the magnetic structure of the star is pure dipolar, or it contains high moments.
Below we show that the interaction of an axionic field with the strong dipolar magnetic field inevitably provides the appearance of quadruple, octopole, etc., moments, thus redistributing the total energy of the magnetic field between its multipolar moments. This means that, in principle, the axionically distorted static magnetic structure is ready for the mild scenario of the restructuring; here we do not describe the corresponding dynamic procedure, but consider the obtained results as a hint for its modeling in future.

The paper is organized as follows. In Sec. II, we recall basic details of the Einstein-Maxwell-axion theory.
In Sec. III, we consider the static model, specify that the gravity field to be of the Reissner-Nordstr\"om type, reduce the set of master equations to the case of magneto-electro-axiono-statics, introduce the appropriate decomposition of the field potentials into the series with respect to the Legendre polynomials, and obtain the equations for the system of radial functions. In Sec. IV we analyze the five moments model and study profiles of the dipolar radial functions.
Section V contains remarks about the structure of high order moments. Section VI is devoted to discussion and conclusions. In the Appendix we discuss the problem of stability of the model.

\section{The Formalism}

\subsection{Action functional}

The typical action functional of the Einstein-Maxwell-axion theory has the form:
$$
S {=}  \int d^4 x \sqrt{{-}g} \left\{\frac
{R{+}2\Lambda}{2\kappa} {-} \frac12  \Psi_{0}^2\left(
\nabla_m\phi \nabla^m\phi {-}V \right) {+}  \right.
$$
\begin{equation}
 \left. {+}\frac{1}{4}F^{mn}\left(F_{mn} {+} \phi F^{*}_{mn} \right) + L^{({\rm matter})} \right\} \,.
\label{action02}
\end{equation}
Here, $g$ is the determinant of the metric tensor $g_{ik}$,
$\nabla_{m}$ is the covariant derivative, $R$ is the Ricci scalar,
$\kappa \equiv 8\pi G$; $G$ is the gravitational
Newtonian coupling constant, $\Lambda$ is the cosmological
constant, and $L^{({\rm matter})}$ is the Lagrangian of the baryonic matter, which forms the solid body of the star. We use the units in which $\hbar$, the Planck constant, and $c$, the speed of
light in vacuum, are chosen to be equal to one. The Maxwell tensor $F_{mn}$ is given by the standard formula
\begin{equation}
F_{mn} \equiv \nabla_m A_{n} - \nabla_n A_{m} \,,
\label{maxtensor}
\end{equation}
where $A_m$ is a potential four-vector of the macroscopic electromagnetic field; $F^{*mn}
\equiv \frac{1}{2} \epsilon^{mnpq}F_{pq}$ is the tensor dual to
$F_{pq}$; $\epsilon^{mnpq} \equiv \frac{1}{\sqrt{-g}} E^{mnpq}$ is
the Levi-Civita tensor, and $E^{mnpq}$ is the skew-symmetric
symbol with $E^{0123}{=}1$. The divergence of the dual Maxwell tensor is equal to zero,
 \begin{equation}
\nabla_{k} F^{*ik} =0 \,, \label{maxtensor2}
\end{equation}
because of the definition (\ref{maxtensor}).
The term $\frac{1}{4}\phi F^{*}_{mn} F^{mn}$ in (\ref{action02}) describes
the pseudoscalar-photon interaction \cite{Ni77}.
We use in this work the symbol $\phi$
for the effective dimensionless pseudoscalar (axion) field; with this dimensionless multiplier $\phi(x)$
two electromagnetic terms in the Lagrangian $\frac{1}{4} F_{mn} F^{mn}$ and $\frac14 \phi(x) F^*_{mn} F^{mn}$  are presented similarly without additional physical parameters \cite{Ni77}.
There is an alternative representation of the second term, which operates with the macroscopic  axion
field, $\Phi$ proportional to this dimensionless quantity
$\Phi {=} \Psi_0 \phi$. Here the auxiliary constant $\Psi_0$ is reciprocal to the coupling constant of the axion-photon-photon interaction  $g_{({\rm A}\gamma \gamma)}$, and  the term $\frac{1}{4}\phi F^{*}_{mn} F^{mn}$ can be rewritten as $\frac{1}{4}g_{({\rm A}\gamma \gamma)}\Phi F^{*}_{mn} F^{mn}$ \cite{Sikivie83}.

\subsection{Master equations}

The standard procedure of variation of the action functional (\ref{action02}) with respect to $A_i$,$\phi$ and $g^{ik}$ yields, respectively, the master equations of the model,
\begin{equation}
\nabla_k \left[F^{ik} {+} \phi  F^{*ik} \right] = 0 \,, \label{eld1}
\end{equation}
\begin{equation}
\nabla_k \nabla^k \phi + \frac12 \frac{\partial V}{\partial \phi} = -
\frac{1}{4\Psi_{0}^2} F^{*}_{mn}F^{mn} \,,
\label{ax1}
\end{equation}
\begin{equation}
R_{ik} - \frac12 R g_{ik} =  \Lambda g_{ik} + \kappa T^{({\rm total})}_{ik} \,.
\label{grav1}
\end{equation}
Here, $T^{({\rm total})}_{ik}$ is the total stress-energy tensors, which contains all the contributions  from the matter, electromagnetic and axion fields.

\section{Static spherically symmetric truncated model}

\subsection{Spacetime symmetry}

We consider the spacetime to be static, spherically symmetric and to be described by the metric
\begin{equation}
ds^2 = N(r)dt^2-\frac{1}{N(r)}dr^2- r^2(d\theta^2 + \sin^2{\theta}d\varphi^2) \,.
\label{metric}
\end{equation}
We assume that the cosmological constant is equal to zero, $\Lambda=0$, and that the axion field inherits partially the symmetry of the spacetime; this means, that not only the metric (\ref{metric}), but the pseudoscalar field $\phi$ also depends neither on time, nor on the azimuthal variable. We postulate that the dipolar, quadruple, etc., contributions of the electromagnetic field and the contribution of the axion field
into the sources of the gravity field  are negligible in comparison with the contributions of the dense matter.
This ansatz means that the metric (\ref{metric}) is assumed to be given, and we search for the pseudoscalar and electromagnetic fields in the gravitational {\it background}. In fact, we assume that the metric is of the Reissner-Nordstr\"om type,
\begin{equation}
N(r) = \left( 1 - \frac{r_{g}}{r} + \frac{r^{2}_{Q}}{r^2} \right) \,,
\label{metric2}
\end{equation}
where $r_{g} = 2GM$ is the Schwarzschild radius, and  $r_{Q} = |Q| \sqrt{G}$ is the radius associated with the charge $Q$ of the object (electric $Q=Q_E$,  magnetic $Q=Q_M$, or hybrid $Q = \sqrt{Q^2_E+Q^2_m}$). The influence of the cosmological constant $\Lambda$ will be studied in a separate paper.

\subsection{Reduced master equations}

\subsubsection{Magneto-electro-statics}

We search for solutions to the electrostatic equations, which inherit the spacetime symmetry. In fact, we consider the electromagnetic field with potentials depending on the radial variable $r$ and the meridional angle $\theta$, which have the form
\begin{equation}
A_i = \delta_i^0 A_0(r, \theta) + \delta_i^{\varphi} A_{\varphi}(r, \theta) \,.
\label {0eld3}
\end{equation}
The Lorentz gauge condition
$$
\nabla_kA^k=0 \ \ \rightarrow \partial_r [r^2 \sin{\theta} A^r] + \partial_{\theta} [r^2 \sin{\theta} A^{\theta}] \equiv  0
$$
is satisfied identically.
In such a model the electrodynamic equations,
\begin{equation}
\partial_k \left (r^2 \sin{\theta} \ g^{im}g^{kn} F_{mn}\right) = - \frac12 E^{ikmn} \partial_k \phi F_{mn}
\label {eld3}
\end{equation}
can be reduced to the pair of equations,
$$
\sin{\theta} \partial_r \left (r^2  \partial_r A_0 \right) + \frac{1}{N} \partial_{\theta} \left (\sin{\theta} \partial_{\theta} A_0 \right)
=
$$
\begin{equation}
= - \partial_{\theta} A_{\varphi} \partial_r \phi +  \partial_r A_{\varphi} \partial_{\theta} \phi \,,
\label {2eld3}
\end{equation}
$$
\frac{1}{\sin{\theta}} \partial_r \left( N \partial_r A_{\varphi}\right) + \frac{1}{r^2}\partial_{\theta} \left(\frac{1}{\sin{\theta}}  \partial_{\theta} A_{\varphi} \right)
=
$$
\begin{equation}
= - \partial_{\theta} A_{0} \partial_r \phi +  \partial_r A_{0} \partial_{\theta} \phi \,.
\label {3eld3}
\end{equation}
Clearly, these equations are coupled when the gradient of the pseudoscalar field is nonvanishing.

\subsubsection{Axiono-statics}

When the pseudoscalar field $\phi$ depends on the radial and meridional variables only, the master equation (\ref{ax1}) can be written as follows:
$$
N \partial^2_r \phi + \frac{1}{r^2} \partial^2_{\theta} \phi + \left[N^{\prime}(r)+\frac{2N}{r} \right]\partial_r \phi + \frac{\cot{\theta}}{r^2} \partial_{\theta}\phi  =
$$
\begin{equation}
= \frac12 \partial_{\phi} V + \frac{1}{4\Psi_{0}^2} F^{*}_{mn}F^{mn} \,.
\label{ax12}
\end{equation}
Our ansatz is that in the dyon environment, the distribution of the pseudoscalar field is guided by the strong  gravitational and electromagnetic fields, and the role of the potential $V(\phi)$ is vanishing. In other words, we work here in the approximation of a massless axion field and consider $V=0$ in all the master equations. Thus, the function $\phi$ satisfies the equation,
$$
N \partial^2_r \phi {+} \left[N^{\prime}(r) {+} \frac{2N}{r} \right] \partial_r \phi {+} \frac{1}{r^2} \left[\partial^2_{\theta}\phi  {+} \cot{\theta}\right] \partial_{\theta}\phi =
$$
\begin{equation}
=
\frac{1}{\Psi_{0}^2 r^2 \sin{\theta}}\left[\partial_{\theta} A_0  \partial_{r} A_{\varphi} -   \partial_{r} A_0 \partial_{\theta} A_{\varphi} \right] \,.
\label{ax123}
\end{equation}

\subsubsection{The known example: axionic dyon}

In order to recover the  model already studied, we consider the functions $\phi$ and  $A_0$ to be the functions of the radial variable only and assume that $A_{\varphi} = \mu (1{-}\cos{\theta})$ does not depend on the radial variable. Then the Eq. (\ref{3eld3}) is satisfied identically;  the Eqs. (\ref{2eld3}) and (\ref{ax123}) convert, respectively,  into
\begin{equation}
\left(r^2A_0^{\prime} {+} \mu \phi \right)^{\prime} =0 \,, \quad \left(r^2 N \phi^{\prime} {+} \frac{\mu}{\Psi^2_0} A_0 \right)^{\prime} =0 \,.
\label{mono1}
\end{equation}
Here and below the prime denote the ordinary derivative with respect to the argument of the function. The equations (\ref{mono1}) coincide with the key equations obtained in \cite{BG2019} for the axionic dyon.
The corresponding interpretation is based on the following procedure. Let us introduce the velocity four-vector $U^i = \delta^i_0 \frac{1}{\sqrt{N}}$ and use the so-called
$(\EuScript{E}, \EuScript{B})$ representation, based on the definition of the four-vectors of the electric and magnetic fields:
\begin{equation}
\EuScript{E}_i \equiv F_{ik} U^k  \,,  \quad \EuScript{B}^i \equiv F^{*ik} U_k     \,.
\label{B109}
\end{equation}
With these definitions for the monopole model we obtain
$$
\EuScript{E}_{0} = 0 \,, \quad \EuScript{E}_{r} = \frac{1}{\sqrt{N}}A^{\prime}_{0} \,,
\quad \EuScript{E}_{\theta} = 0 \,, \quad
\EuScript{B}^{0} = 0 \,,
$$
\begin{equation}
\EuScript{B}^{r} =  - \sqrt{N} \frac{\mu}{r^{2}} \,, \quad
\EuScript{B}^{\theta} = 0 \,, \quad \EuScript{B}^{\varphi} = 0  \,.
\label{E&B22}
\end{equation}
In other words, the model describes configuration with pure radial electric and magnetic fields
\begin{equation}
\EuScript{B}_{(\rm rad)} \equiv \sqrt{{-}\EuScript{B}^{r}\EuScript{B}_{r}} {=} \frac{\mu}{r^2} \,, \
\EuScript{E}_{(\rm rad)} \equiv \sqrt{{-}\EuScript{E}^{r}\EuScript{E}_{r}} {=} A_0^{\prime} \,.
\label{mono4}
\end{equation}
Let us emphasize that the model of the axionic dyon is unique, since this model admits the set of exact solutions for the axion, electric, and magnetic fields with a pure monopole structure.
When the magnetic field has, at least, a dipole component, the corresponding axion and electric fields also depend on the angular variables (on the variable $\theta$ in our truncated model), and all the higher moments: quadruple, octupole, etc., inevitably appear.

\subsection{Angular structure of the field potentials}

We search for the electromagnetic potentials and pseudoscalar field in the following form:
\begin{equation}
A_{0}(r,\theta) = \sum_{n=0}^{\infty} F_{n}(r)P_{n}(\cos\theta) \,,
\label{0radial0}
\end{equation}
$$
A_{\varphi}(r,\theta) = - G_0(r) - G^{*}_0(r) \cos{\theta} -
$$
\begin{equation}
- \sum_{m=1}^{\infty} G_{m}(r)\sin\theta \frac{d}{d\theta} P_{m}(\cos\theta) \,,
\label{0radial1}
\end{equation}
\begin{equation}
\phi(r,\theta) = \sum_{k=0}^{\infty} \psi_k(r) P_k(\cos{\theta}) \,.
\label{ax122}
\end{equation}
Here, $P_k(\cos{\theta})$ are the Legendre polynomials;  below, we use also the definition $z=\cos{\theta}$. The quantities $F_{n}(r)$, $G_{m}(r)$ and $\psi_k(r)$ are the radial functions to be found.
Let us mention that the first and second terms in (\ref{0radial1}) contain the standard Legendre polynomials $P_0=1$ and $P_1 = \cos{\theta}$, while other terms contain the adjoint Legendre polynomials $P^{(1)}_m = \sqrt{1{-}z^2} \ \frac{dP_m}{dz}= - \frac{d}{d\theta} P_{m}(\cos\theta)$. This decomposition is motivated by the following arguments.
In the $(\EuScript{E}, \EuScript{B})$ representation the physical components of the electric and magnetic fields, given by the potentials (\ref{0radial0}) and (\ref{0radial1},) can be written as follows:
$$
\EuScript{E}_{r} = \frac{1}{\sqrt{N}} \sum_{n=0}^{\infty} F^{\prime}_{n}(r)P_{n}(z) \,,
$$
$$
\EuScript{E}_{\theta} = - \frac{1}{\sqrt{N}} \sum_{n=0}^{\infty} F_{n}(r)\sqrt{1-z^2} \ P^{\prime}_{n}(z)\,,
$$
$$
\EuScript{E}_{0} = 0 \,, \quad  \EuScript{E}_{\varphi} = 0 \,, \quad \EuScript{B}^{0} = 0 \,, \quad
\EuScript{B}^{\varphi} = 0 \,,
$$
$$
\quad \EuScript{B}^{r} =  \frac{\sqrt{N}}{r^{2}}\left[ G^{*}_0(r)  {+}  \sum_{m=1}^{\infty} m(m+1)G_{m}(r) P_{m}(z) \right] \,,
$$
$$
\EuScript{B}^{\theta} =  \frac{\sqrt{N}}{r^{2}}\left[- \frac{G^{\prime}_0(r) + z G_0^{* \prime}(r)}{\sqrt{1-z^2}}  + \right.
$$
\begin{equation}
\left. + \sum_{m=1}^{\infty} G^{\prime}_{m}(r) \sqrt{1-z^2} \ \frac{d}{dz} P_{m}(z) \right] \,.
\label{E&B223}
\end{equation}
When one deals with  a magnetic monopole with the charge $\mu$ and radial magnetic field only, one can recover the classical formulas by putting
\begin{equation}
-G_0 =  G_0^{*}= \mu = const \,,  \quad G_1=G_2=...=0 \,.
\label{dip0}
\end{equation}
When we deal with a magnetic dipole, we keep in mind the classical formulas
\begin{equation}
\EuScript{B}_{(\rm rad)} = - \frac{2\nu}{r^3} \cos{\theta}\,, \quad \EuScript{B}_{(\rm merid)} = \frac{\nu}{r^3} \sin{\theta} \,.
\label{dip1}
\end{equation}
Clearly, we can recover these formulas in the limit $N \to 1$, if to put
\begin{equation}
G_0 =  G_0^{*} = 0 \,, \quad G_1 = - \frac{\nu}{r} \,, \quad G_2 =G_3= ... =0 \,.
\label{E&B228}
\end{equation}
These two examples help us to properly represent the decomposition (\ref{0radial1}).

\subsection{Master equations for the radial functions}

Searching for the radial functions $F_n(r)$, $G_m(r)$, and $\psi_k(r)$, we have to work with the Legendre polynomials, and to take into account the following three their cardinal properties: first, the basic equations for $P_n(z)$,
\begin{equation}
\frac{d}{d z} \left[ (1-z^2) \frac{d}{dz} P_{n} \right] + n(n+1) P_n = 0 \,,
\label{000angle}
\end{equation}
second, the orthogonality-normalization conditions for the standard Legendre polynomials $P_n(z)$ and adjoint Legendre polynomials $P^{(1)}_n(z)$,
\begin{equation}
\int_{-1}^{1}dz P_m(z)P_n(z) = \frac{2}{2n+1} \delta^m_{n} \,,
\label{00angle}
\end{equation}
\begin{equation}
 \int_{-1}^{1}dz P^{(1)}_m(z)P^{(1)}_n(z) = \frac{2m(m+1)}{2m+1} \ \delta^m_{n} \,,
\label{00angle1}
\end{equation}
third, two consequences from the recurrent formulas,
\begin{equation}
(1-z^2)\frac{dP_m}{dz}=\frac{m(m+1)}{2m+1}\left(P_{m-1}-P_{m+1} \right) \,,
\label{P17}
\end{equation}
\begin{equation}
(2n+1)zP_n(z)= (n+1) P_{n+1} + n P_{n-1} \,.
\label{P18}
\end{equation}
{\it As the first step}, we put $A_{\varphi}$ given by (\ref{0radial1}) into the Eq. (\ref{3eld3}), and see
immediately, that the necessary conditions for the compatibility of the angular decomposition require
\begin{equation}
\left( NG_0^{\prime}\right)^{\prime} = 0 \,, \quad \left( NG_0^{*\prime}\right)^{\prime} = 0 \,.
\label{00*}
\end{equation}
The corresponding general solutions
\begin{equation}
G_0 = K_2 + K_1 \int \frac{dr}{N(r)} \,, \quad G^*_0 = K^*_2 + K^*_1 \int \frac{dr}{N(r)}
\label{10*}
\end{equation}
show that the monopole terms in (\ref{0radial1}) with $G_0$ and $G^*_0$ happen to be decoupled from other multipoles. In other words, if there was no magnetic charge in the system under consideration, the monopole - type magnetic moment  can not appear because of interaction with dipole, quadruple, etc., moments. This is the argument for us to put below $G_0=0=G_0^*$, thus considering only the new magnetic structures in comparison with the results of \cite{BG2019}, describing the monopole type solutions.

{\it As the second step}, we  multiply (\ref{0radial1}) by $\frac{dP_m}{d\theta}$ ($m=1,2,...$), integrate with respect to $z=\cos{\theta}$, and obtain the system of equations for $G_m(r)$ ($m=1,2,...$),
\begin{equation}
\left(N G_m^{\prime} \right)^{\prime} {-} \frac{m(m{+}1)}{r^2} G_m {=} \sum_{n=0}^{\infty} \sum_{k=0}^{\infty} h^{(3)}_{mkn}\left[F_n \psi^{\prime}_k {-}   \psi_n F^{\prime}_k \right] \,.
\label{9angle32}
\end{equation}
{\it As the third step}, we take the Eq. (\ref{2eld3}) with $A_0$ from (\ref{0radial0}) multiplied by $\frac{P_s(z)}{\sqrt{1-z^2}}$, and after integration
with respect to $z$, keeping in mind (\ref{000angle}),  we obtain
$$
\left(r^2 F^{\prime}_n \right)^{\prime} - \frac{n(n+1)}{N} F_n =
 $$
 \begin{equation}
 = - \sum_{k=0}^{\infty} \sum_{m=1}^{\infty} \left[h^{(1)}_{mkn}G_m \psi_k^{\prime} +   h^{(2)}_{mkn} \psi_k G_m^{\prime} \right]  \,,
\label{angle31}
\end{equation}
where $n=0,1,...$.

{\it As the fourth step}, we consider the equations for the radial functions $\psi_k(r)$; for this purpose we put the decompositions (\ref{0radial0}), (\ref{0radial1}), and (\ref{ax122}) into (\ref{ax123}), and integrate this equation multiplied by $P_s(z)$  with respect to $z$. This standard procedure yields the equation
$$
\Psi^2_0 \left[\left(r^2 N \psi_k^{\prime}\right)^{\prime}- k(k+1)\psi_k \right] =
$$
\begin{equation}
= - \sum_{n=0}^{\infty} \sum_{m=1}^{\infty}\left[F_n^{\prime} G_m h^{(1)}_{mnk} + F_n G_m^{\prime} h^{(2)}_{mnk} \right] \,.
\label{P19}
\end{equation}
In the key equations (\ref{9angle32}), (\ref{angle31}), and (\ref{P19}) the following auxiliary coefficients are introduced:
\begin{equation}
h^{(1)}_{mkn} \equiv \frac{1}{2}(2n{+}1)m(m{+}1)\int_{{-}1}^1 dz  P_m(z) P_k(z) P_n(z) \,,
\label{angle35}
\end{equation}
\begin{equation}
h^{(2)}_{mkn} \equiv \frac{1}{2}(2n{+}1) \int_{{-}1}^1 dz (1{-}z^2) P^{\prime}_m(z)  P^{\prime}_k(z) P_n(z) \,.
\label{angle36}
\end{equation}
\begin{equation}
h^{(3)}_{mkn} \equiv \frac{2m{+}1}{2m(m{+}1)}\int_{{-}1}^1 dz (1{-}z^2) P^{\prime}_m(z) P_k(z) P^{\prime}_n(z) \,.
\label{angle37}
\end{equation}
We  see that all the right-hand sides of the key equations (\ref{9angle32}), (\ref{angle31}), and (\ref{P19}) contain quadratic cross-terms. Indeed, the equations for the magnetic radial functions $G_m$ contain the sources formed by the products of axionic and electric contributions; the equations for the electric radial functions $F_n$ contain the sources formed by the products of axionic and magnetic contributions; similarly, the equations for the axionic radial functions $\psi_k$ contain the sources formed by the products of electric and magnetic contributions.
In order to analyze this whole coupled system of equations we have to discuss the scheme of calculation of the auxiliary coefficients $h^{(1)}_{mkn}$, $h^{(2)}_{mkn}$, and $h^{(3)}_{mkn}$.

\subsection{Scheme of calculation of auxiliary coefficients}

When we calculate the integrals (\ref{angle35}), we have to keep in mind that the product of the Legendre polynomials $P_m P_n $ is a polynomial of the order $m+n$, and it can be decomposed into a series with respect to $P_l$ as follows (see, e.g., \cite{Hobson}):
\begin{equation}
P_m(z) P_n(z) = \sum_{l=0} \alpha_{m+n-2l} P_{m+n-2l} \,,
\label{h11}
\end{equation}
\begin{equation}
 \alpha_{p+q} = \frac{[(p+q)!]^2 (2p)! (2q)!}{(2p+2q)! (p!q!)^2} \,.
\label{h12}
\end{equation}
If $m>n$, the last term in this decomposition contains $P_{m{-}n}$, the Legendre polynomial with the index $m{-}n$.

\subsubsection{Calculation of the coefficients $h^{(1)}_{mkn}$}

The decomposition (\ref{h11}) allows us to calculate directly the integrals (\ref{angle35}), which contain the products of three Legendre polynomials. Let us mention that these integrals are equal to zero, when the sum of any two indices is less than the third one. When, e.g., $m+n \geq k$, $m \geq n$ and $m+n+k$ is the even number, one obtains (see \cite{Hobson})
$$
\int_{-1}^1 dz  P_m(z) P_k(z) P_n(z) =
$$
$$
= \frac{2}{(m+n+k+1)}  \times \frac{2 \cdot 4 ... (m+n+k)}{1 \cdot 3 ... (m+n+k-1)}  \times
$$
$$
\times \frac{1 \cdot 3 ... (m+n-k-1)}{2 \cdot 4 ... (m+n-k)}  \times \frac{1 \cdot 3 ... (m+k-n-1)}{2 \cdot 4 ... (m+k-n)}
$$
\begin{equation}
\times \frac{1 \cdot 3 ... (k+n-m-1)}{2 \cdot 4 ... (k+n-m)}  \,.
\label{anglez35}
\end{equation}
In other words, for the coefficients $h^{(1)}_{mkn}$ we can use the known formulas.

\subsubsection{Calculation of the coefficients $h^{(2)}_{mkn}$}

When we consider the integrals (\ref{angle36}), we use the formula (\ref{P17}) and obtain the integrals
$$
h^{(2)}_{mkn}= \frac{(2n{+}1)m (m{+}1)}{2(2m{+}1)} \times
$$
\begin{equation}
\times \int_{{-}1}^1 dz \left(P_{m{-}1}{-}P_{m{+}1} \right)  P^{\prime}_k(z) P_n(z) \,.
\label{P256}
\end{equation}
This means that now we have to calculate the integrals of the following type:
\begin{equation}
{\cal J}_{pql} = {\cal J}_{qpl} \equiv \int_{-1}^1 dz P_p(z) P_q(z) P^{\prime}_l(z) \,.
\label{P2}
\end{equation}
Keeping in mind the evident formula,
\begin{equation}
{\cal J}_{pql} {+} {\cal J}_{qlp} {+} {\cal J}_{plq} = \int_{-1}^1 d \left[P_p P_q P_l\right] =  1{-}({-}1)^{p{+}q{+}l} \,,
\label{P3}
\end{equation}
and computing the terms with $l=0$, $l=1$, etc., we can ease this task essentially. For instance, we obtain directly the following two consequences:
$$
{\cal J}_{pq0}=0  \Rightarrow
$$
\begin{equation}
2{\cal J}_{q0p} = 1-(-1)^{p+q} \,, \quad {\cal J}_{q0q} =0 \,, \quad {\cal J}_{q+1, 0,q} = 1 \,,
\label{P47}
\end{equation}
$$
{\cal J}_{pq1}=\delta^p_q \frac{2}{2p+1} \Rightarrow
$$
\begin{equation}
{\cal J}_{q1q} = \frac{2q}{2q+1} \,, \quad {\cal J}_{q+1,0q} =0 \,.
\label{P48}
\end{equation}

\subsubsection{Calculation of the coefficients $h^{(3)}_{mkn}$, and results for some sets of numbers $m,k,n$}

The integral in (\ref{angle37}) can be obtained from the integral in (\ref{angle36}) using the replacement $k \leftarrow \rightarrow n$.
As an example, using this technology, we obtain the coefficients of the first type,
$$
h^{(1)}_{0kn} = 0 \,,  h^{(1)}_{m0n} = m(m{+}1) \ \delta^m_n \,, h^{(1)}_{mk0} = \frac{m(m{+}1)}{2m{+}1}  \delta^m_k \,,
$$
$$
h^{(1)}_{m1n} = \frac{m(m+1)}{(2m+1)} \left[(n+1)\delta^m_{n+1} + n \delta^m_{n-1} \right] \,,
$$
$$
h^{(1)}_{mk1} = \frac{3m(m+1)}{(2m+1)(2k+1)} \left[(k+1)\delta^m_{k+1} + k \delta^m_{k-1} \right] \,,
$$
\begin{equation}
h^{(1)}_{1kn} = \frac{2}{(2k+1)} \left[(n+1)\delta^k_{n+1} + n \delta^k_{n-1} \right] \,,
\label{P491}
\end{equation}
the coefficients of the second type,
$$
h^{(2)}_{0kn} = h^{(2)}_{m0n} = 0 \,, \quad h^{(2)}_{mk0} = \delta^m_k \frac{m(m+1)}{2m+1} \,,
$$
$$
h^{(2)}_{1kn}= \frac{k(k+1)}{(2k+1)} \left[\delta^n_{k-1} - \delta^n_{k+1} \right] \,,
$$
$$
 h^{(2)}_{m1n} = \frac{m(m+1)}{(2m+1)} \left[\delta^n_{m-1} - \delta^n_{m+1} \right] \,,
$$
and the coefficients of the third type,
$$
h^{(3)}_{0kn} = 0 \,, \quad h^{(3)}_{m0n} = \delta^m_n \,, \quad
h^{(3)}_{mn0} = 0 \,,
$$
$$
h^{(3)}_{mn1} = \frac{1}{2n+1} \left(\delta^n_{m-1}-\delta^n_{m+1} \right) \,,
$$
$$
h^{(3)}_{11n} = \frac35 \delta^n_2 \,, \quad h^{(3)}_{21n} = \frac47 \delta^n_3 + \frac13 \delta^n_1\,, \quad h^{(3)}_{m00} = 0 \,,
$$
\begin{equation}
h^{(3)}_{m01} = \delta^m_1 \,, \quad h^{(3)}_{m10} =  0 \,, \quad h^{(3)}_{m11} = \frac{2}{m(m{+}1)}\delta^m_2  \,,
\label{P49}
\end{equation}
which will be necessary below.

\section{Five moments model}

In this section we consider the model, which is based on the interplay of five radial functions: $F_0(r)$, $\psi_0(r)$, $F_1(r)$, $G_1(r)$, $\psi_1(r)$. Two first functions can be indicated as the monopole type ones; we assume that $G_0=0$; i.e., the magnetic monopole moment is absent. Next three functions from this quintet are of the dipole type. In fact, this five-moments model has the so-called $2+3$ construction, i.e., the equations for the monopole and dipole functions can be decoupled. Let us discuss this idea.

\subsection{Master equations for the monopole moments of the electric and axion fields}

Let us consider the Eq. (\ref{angle31}) for $n=0$ and Eq. (\ref{P19}) for $k=0$ with the coefficients $h^{(1)}_{mkn}$ $h^{(2)}_{mkn}$, presented in the previous section. We obtain the master equations for the radial functions $F_0(r)$ and $\psi_0(r)$, which describe the monopole moments of the electric and pseudoscalar fields, respectively,
\begin{equation}
\left(r^2 F_0^{\prime} \right)^{\prime} = - \sum_{m=1}^{\infty} \frac{m(m+1)}{(2m+1)} \left(G_m \psi_m \right)^{\prime} \,,
\label{L1}
\end{equation}
\begin{equation}
\left(r^2 N \psi_0^{\prime} \right)^{\prime} = - \frac{1}{\Psi^2_0}\sum_{m=1}^{\infty} \frac{m(m+1)}{(2m+1)} \left(F_m G_m \right)^{\prime} \,.
\label{L2}
\end{equation}
Integration of (\ref{L1}) gives immediately
\begin{equation}
F_0^{\prime}(r)  = - \frac{Q}{r^2} - \frac{1}{r^2} \sum_{m=1}^{\infty} \frac{m(m+1)}{(2m+1)} G_m(r)  \ \psi_m(r)   \,,
\label{L11}
\end{equation}
where the constant of integration $Q$ can be interpreted standardly as the total electric charge of the magnetic star, and the term $\frac{Q}{r}$ relates to the Coulombian part of the electric field.
Integration of (\ref{L2}) yields
\begin{equation}
\psi_0^{\prime}(r)  = \frac{K}{r^2 N } - \frac{1}{\Psi^2_0 r^2 N} \sum_{m=1}^{\infty} \frac{m(m+1)}{(2m+1)} F_m G_m  \
\,,
\label{L21}
\end{equation}
where the constant $K = \lim_{r \to \infty} \left(r^2 N \psi_0^{\prime} \right)$ can be interpreted as the axionic charge \cite{LW1991,BG2019}. The formula (\ref{L11}) shows that the electric field of the magnetic star can be intrinsic (with the total charge $Q$) and the axionically induced; clearly, all the radial functions of the axion field contribute to the electric field of the monopole type. The formula (\ref{L21}) demonstrates that the axionic field can also be intrinsic (with the charge $K$) and induced by the interaction with the electromagnetic field as well as all of the radial functions $G_m$ and $F_m$ can contribute to the axionic field of the monopole type.

\subsection{Master equations for the dipole moments of the electric, magnetic and axion fields}

In the case when we can neglect the quadruple, etc. moments of the magnetic, electric and axion field, we can consider the following model system of equations:
\begin{equation}
 \left(r^2 F^{\prime}_1 \right)^{\prime} - \frac{2}{N} F_1 =  -  2 G_1 \psi^{\prime}_0   \,,
\label{7angle313}
\end{equation}
\begin{equation}
\left(N G_1^{\prime} \right)^{\prime} - \frac{2}{r^2} G_1 = \psi_0^{\prime}F_1 - \psi_1 F_{0}^{\prime}  \,,
\label{angle329}
\end{equation}
\begin{equation}
\left(r^2 N \psi_1^{\prime}\right)^{\prime}- 2\psi_1  =  - \frac{2}{\Psi^2_0 }F_0^{\prime} G_1  \,.
\label{1P19}
\end{equation}
Then we replace $\psi_0^{\prime}$ and $F_{0}^{\prime}$ using (\ref{L11}) and (\ref{L21}) reduced, respectively, to
\begin{equation}
F_0^{\prime}(r)  = - \frac{Q}{r^2} - \frac{2}{3 r^2} G_1  \psi_1 \,,
\label{L111}
\end{equation}
\begin{equation}
\psi_0^{\prime}(r)  = \frac{K}{r^2 N } - \frac{2}{3\Psi^2_0 r^2 N} G_1 F_1  \,,
\label{L1141}
\end{equation}
thus obtaining three coupled nonlinear equations for the dipole-type radial functions
\begin{equation}
 \left(r^2 F^{\prime}_1 \right)^{\prime} - \frac{2}{N} F_1 \left[1+   \frac{2}{3\Psi^2_0 r^2}  G_1^2  \right]=  -   \frac{2K}{r^2 N } G_1
    \,,
\label{key1}
\end{equation}
\begin{equation}
\left(N G_1^{\prime} \right)^{\prime} {-} \frac{2}{r^2} G_1\left[1{-} \frac{1}{3\Psi^2_0 N}  F^2_1 {+}  \frac{1}{3}  \psi_1^2  \right]   = \frac{K}{r^2 N }F_1  {+} \frac{Q}{r^2}\psi_1    \,,
\label{key2}
\end{equation}
\begin{equation}
\left(r^2 N \psi_1^{\prime}\right)^{\prime}- 2\psi_1 \left[1+ \frac{2}{3\Psi^2_0  r^2} G^2_1  \right]  =   \frac{2Q}{\Psi^2_0 r^2} G_1
\,.
\label{key3}
\end{equation}
When we describe the magnetic star with an extremely high magnetic field, we can use the following hierarchical approach to solve the coupled system of the nonlinear equations (\ref{key1}), (\ref{key2}), and (\ref{key3}). We consider the magnetic radial function $G_1$ to be decoupled, since $F_1$ and $\psi_1$ add to this function the linear and quadratic terms in $\frac{1}{\Psi^2_0}$. In other words, we extract from the coupled system of equations, the equation for $G_1$ in the truncated form,
\begin{equation}
\left(N G_1^{\prime} \right)^{\prime} - \frac{2}{r^2} G_1  = 0   \,.
\label{key4}
\end{equation}
Then after the analysis of this equation, we study the remaining pair of equations for the pair of functions $F_1(r)$ and $\psi_1(r)$ [see (\ref{key1}) and (\ref{key3}), respectively].
Let us mention that the Eq. (\ref{key4}) can also be obtained as the exact consequence of (\ref{key2}), when $K=0$ and $Q=0$, since for these parameters, the Eqs. (\ref{key1}) and (\ref{key3}) admit the trivial solutions  $\psi_1=0$ and $F_1=0$.

\subsection{Profile of the radial function $G_1(r)$}

\subsubsection{Key equation}

When we analyze the Eq. (\ref{key4}) with the Reissner-Nordstr\"om metric function $N(r)= 1- \frac{r_g}{r}+ \frac{r^2_{Q}}{r^2}$, we use  the dimensionless variable $x$ and parameter $a$, defined as follows:
\begin{equation}
x = \frac{r}{r_{+}} \,, \quad a \equiv \frac{r_g}{r_{+}} -1 = \frac{r_{-}}{r_{+}} < 1 \,,
\label{au1}
\end{equation}
\begin{equation}
\quad r_{\pm} = \frac12 r_g \left[1 \pm \sqrt{1-\frac{4r^2_{Q}}{r^2_g}} \right] \,.
\label{au11}
\end{equation}
When $a \neq 0$ and $a \neq 1$, in these terms, the Eq. (\ref{key4}) takes the form,
\begin{equation}
G_1^{\prime \prime} + G_1^{\prime} \left[ \frac{1}{x{-}a} {+} \frac{1}{x{-}1} {-} \frac{2}{x} \right] -  G_1 \frac{2}{(x{-}a)(x{-}1)}  = 0   \,.
\label{D11}
\end{equation}
Let us compare (\ref{D11}) with the Heun equation \cite{Heun1,Heun2}, which is a particular case of the known Fuchs equation \cite{Ince,Poole},
\begin{equation}
Y^{\prime \prime} + Y^{\prime} \left[ \frac{\epsilon}{x{-}a} {+} \frac{\delta}{x{-}1} {+} \frac{\gamma}{x} \right] +  Y \frac{\alpha \beta x {-} q}{(x{-}a)(x{-}1)x}  = 0   \,.
\label{Heun}
\end{equation}
One can conclude that the radial magnetic function $G_1(x)$ satisfies the Heun equation (\ref{Heun}) with the parameters
\begin{equation}
\epsilon=1 \,, \quad \delta = 1 \,, \quad \gamma = -2 \,, \quad \alpha \beta =-2 \,, \quad q = 0  \,.
\label{Heun1}
\end{equation}
Taking into account that the Heun parameters are connected by the linear relationship $\epsilon + \gamma + \delta = \alpha + \beta + 1$, which guarantees that infinity is a regular point, we
can choose $\alpha = 1$ and $\beta=-2$, and can specify the solution as follows:
\begin{equation}
G_1 = Y(x,a,1,1,-2,1,-2,0) \,.
\label{Heun14}
\end{equation}

\subsubsection{Two special cases}

The solutions to the key equation (\ref{D11}) form a one-parameter family, $a$ being the guiding parameter, which belongs to the interval $0\leq a \leq 1$; we distinguish two special values of this parameter $a=0$ and $a=1$, which mark the edges of the mentioned interval.

\vspace{2mm}
\noindent
1. $a=0$.

\noindent
This case relates to the model with a vanishing electric charge, $Q=0$. We see now that $r_{+}=r_g$, $r_{-}=0$, $N(r)=1-\frac{r_g}{r}$, and the Eq. (\ref{D11}) converts into
\begin{equation}
x(x-1)G_1^{\prime \prime} + G_1^{\prime} - 2 G_1  = 0   \,.
\label{D19}
\end{equation}
Comparing this equation with the hypergeometric equation (see, e.g., \cite{Bateman}),
\begin{equation}
\zeta(1{-}\zeta)U^{\prime \prime}(\zeta) {+} \left[\gamma{-}(\alpha{+}\beta{+}1) \zeta \right] U^{\prime}(\zeta) {-} \alpha \beta U = 0\,,
\label{G1}
\end{equation}
we conclude that $G_1$ satisfies the hypergeometric equation with the parameters $\alpha=1$, $\beta=-2$, $\gamma =-1$; i.e.,
\begin{equation}
G_1 = F(1,-2,-1,x)\,,
\label{psi17}
\end{equation}
where the letter $F$ denotes the hypergeometric function.

\vspace{2mm}
\noindent
2. $a=1$.

\noindent
Now we deal with the extremal case, since when  $a=1$ the inner and outer horizons coincide:
\begin{equation}
a=1 \ \ \rightarrow   2r_{Q}= r_g \ \ \rightarrow    r_{+} = r_{-} = \frac12 r_g \,.
\label{w2}
\end{equation}
The equation for $G_1$ takes now the form
\begin{equation}
G_1^{\prime \prime} + 2G_1^{\prime} \left[ \frac{1}{x-1} - \frac{1}{x} \right] -  G_1 \frac{2}{(x-1)^2}  = 0   \,,
\label{D19i}
\end{equation}
and the corresponding solution
\begin{equation}
G_1 = Y(x,1,\epsilon,2-\epsilon,-2,1,-2,0)
\label{Heun142}
\end{equation}
has one arbitrary hidden parameter $\epsilon$; as it was previously, $Y$ is the corresponding Heun's function.

\subsubsection{Asymptotic behavior and the scheme of numerical analysis}

In the far zone, when $r \to \infty$ and $N(r \to \infty) \to 1$, we can rewrite the key equation (\ref{D11}) as follows:
\begin{equation}
x^2 G_1^{\prime \prime} + (a+1)G_1^{\prime} -  2G_1  = 0 \,.
\label{2D13}
\end{equation}
The regular at infinity asymptotic solution to this equation is of the form
\begin{equation}
G_1(x) \to \frac{G^*_1}{x}\left[1+ \frac{(a+1)}{4x}+ ... \right]  \,.
\label{D13}
\end{equation}
In other words, the Eq. (\ref{key4}) admits that an one-parameter family of solutions, which belong to the class of the Heun functions $Y(x,a,1,1,-2,1,-2,0)$, are regular and have Coulombian-type behavior at infinity.
For simplification of the numerical calculations it is convenient to introduce the dimensionless function ${\cal M}(x,a) \equiv \frac{G_1(x)}{G_1(R_0)}$, where the $R_0>r_{+}$ is the radius of the magnetic star solid body. According to (\ref{E&B22}), the parameter $G_1(R_0)$ can be associated with the radial magnetic field on the north pole of the magnetic star ($\theta=0$) as follows:
\begin{equation}
G_1(R_0) = \frac{1}{2} R^2_0 {\cal B}_{(\rm rad)}(R_0) \,.
\label{GR}
\end{equation}
Respectively, the profile ${\cal M}(x,a)$ satisfies the Eq. (\ref{D11}) with the replacement $G_1 \to {\cal M}$. Clearly, the asymptotic behavior of the function ${\cal M}(x,a)$ is given by the analytical formula $ {\cal M}(x,a)\to \frac{{\cal M}_{\infty}}{x}\left[1+ \frac{(a+1)}{4x}+ ... \right]$, and we focus on the behavior of the profile in the near zone, the size of which is about several radii of the magnetic star. In Fig. 1 we illustrated the behavior of the profiles ${\cal M}(x,a)$ in the near zone for three values of the parameter $a$.

\begin{figure}
	\includegraphics[width=85mm,height=70mm]{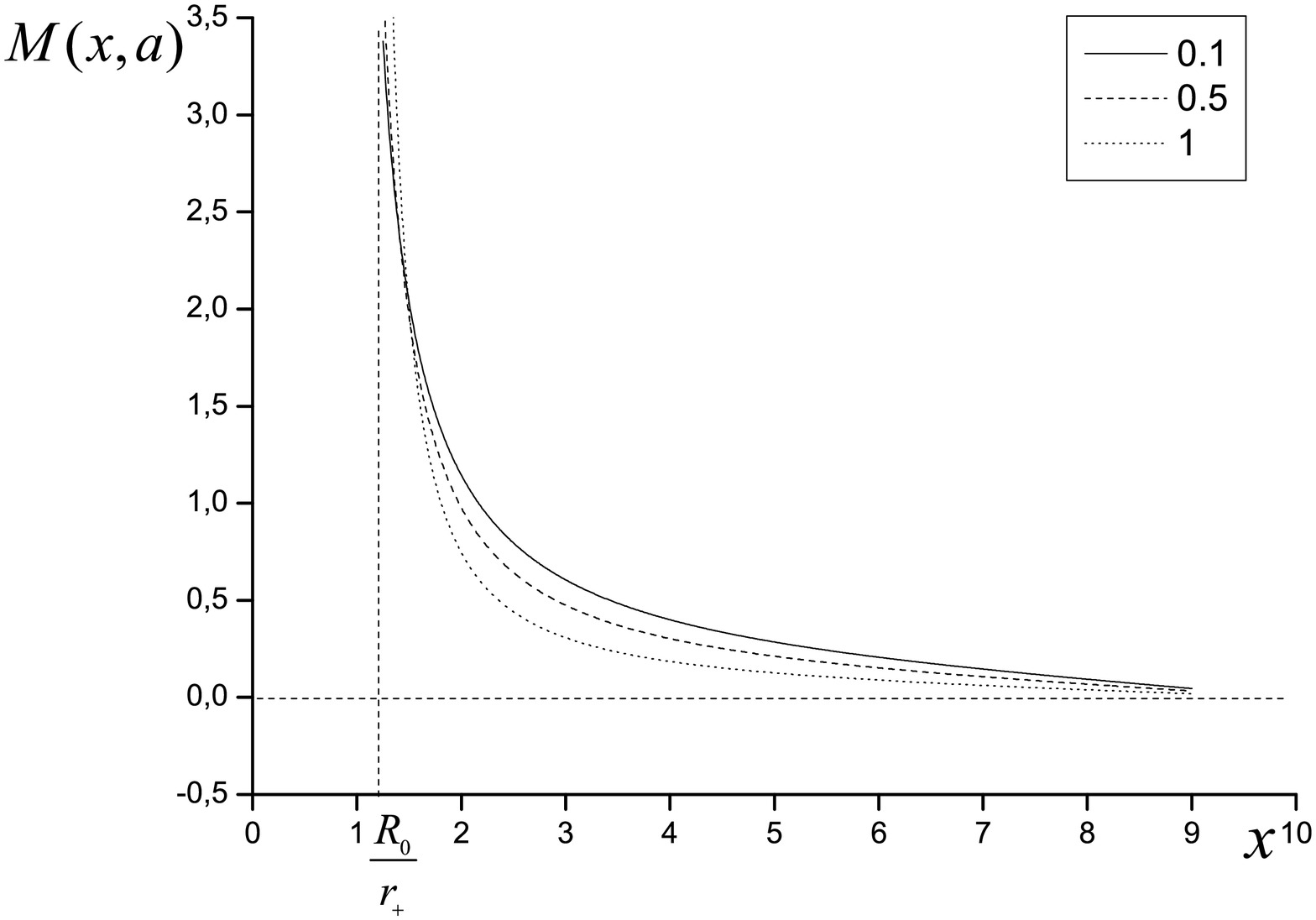}
	\caption{Typical  profiles of the reduced dipolar magnetic radial function ${\cal M}(x,a){=}\frac{G_1(x)}{G_1(R_0)}$ in the near zone presented for three values of the guiding parameter $a$ ($0\leq a\leq 1$). The value $x{=}1$ relates to the position of the outer horizon. The curves start from the points with $x{=}\frac{R_0}{r_{{+}}}$, which correspond to the surface of the star solid body.}
\end{figure}

\subsection{Profile of the dipole type axion field}

The equation for the dipole type axion field (\ref{key3}) can be rewritten in the form,
$$
\psi_1^{\prime \prime} {+} \psi_1^{\prime} \left(\frac{1}{x{-}a} {+} \frac{1}{x{-}1}\right) {-} \frac{2 \psi_1}{(x{-}a)(x{-}1)} \left[1 {+} \omega  \frac{{\cal M}^2(x)}{x^2} \right] =
 $$
 \begin{equation}
 =\frac{\Omega {\cal M}(x)}{x^2(x-a)(x-1)} \,,
\label{GR2key}
\end{equation}
where two auxiliary parameters are introduced,
\begin{equation}
\omega \equiv \frac{R_0^4 {\cal B}^2_{(\rm rad)}(R_0)}{6 r^2_{+} \Psi^2_0} \,, \quad   \Omega \equiv \frac{Q R_0^2 {\cal B}_{(\rm rad)}(R_0)}{ r^2_{+} \Psi^2_0} \,.
\label{GR23}
\end{equation}
Far from the magnetic star surface ($x \to \infty$) the function ${\cal M}(x)$ behaves as ${\cal M}(x) \to \frac{{\cal M}_{\infty}}{x}$, so that asymptotically the key equation (\ref{GR2key}) behaves as the inhomogeneous Euler's equation,
\begin{equation}
x^2 \psi_1^{\prime \prime} + 2x \ \psi_1^{\prime} - 2 \psi_1  =  \Omega {\cal M}_{\infty} x^{-3} \,,
\label{GR2inf}
\end{equation}
whose general solution is
\begin{equation}
 \psi_1(x) = C_1 x + C_2 x^{-2} +  \frac14 \Omega {\cal M}_{\infty} x^{-3} \,.
\label{GR2inf2}
\end{equation}
Clearly, for the asymptotic solutions, we have to put $C_1=0$ and have to claim that the electromagnetically induced axionic dipole type radial function tends to zero asymptotically as $\psi_1 \propto x^{-3}$ [see the last term in (\ref{GR2inf2})].
In the near zone, the profiles of the radial function $\psi_1(x)$ depend essentially on three parameters $a$, $\omega$, $\Omega$. We distinguish two formal cases: $Q=0$ and $Q \neq 0$. The first case relates to the vanishing third parameter $\Omega=0$; the illustrations of the behavior of a two-parameter family of the solutions in the near zone are presented in Fig. 2. When $Q \neq 0$ and thus $\Omega \neq 0$, the behavior of the three-parameter family of solutions is illustrated in Fig. 3.

\begin{figure}
	\includegraphics[width=85mm,height=70mm]{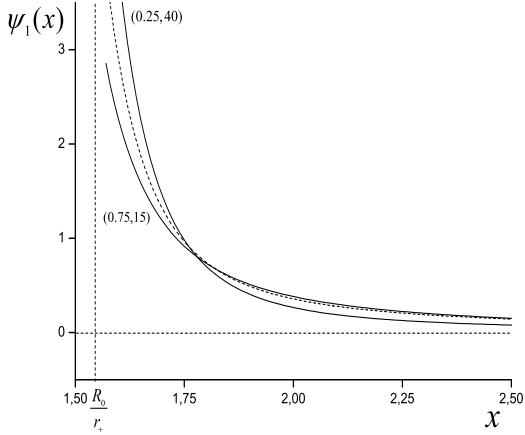}
	\caption{Typical profiles of the dipolar axionic radial function ${\psi_1(x)}$ in the near zone for the case, when the total electric charge of the star and thus the guiding parameter $\Omega$ are vanishing, $Q{=}0$, $\Omega{=}0$.  The values of the guiding parameters $(a,\omega)$ are indicated in parentheses near two curves; the third (dotted) line relates to the special value $a{=}1$ ($\omega{=}10$). }
\end{figure}

\begin{figure}
	\includegraphics[width=85mm,height=70mm]{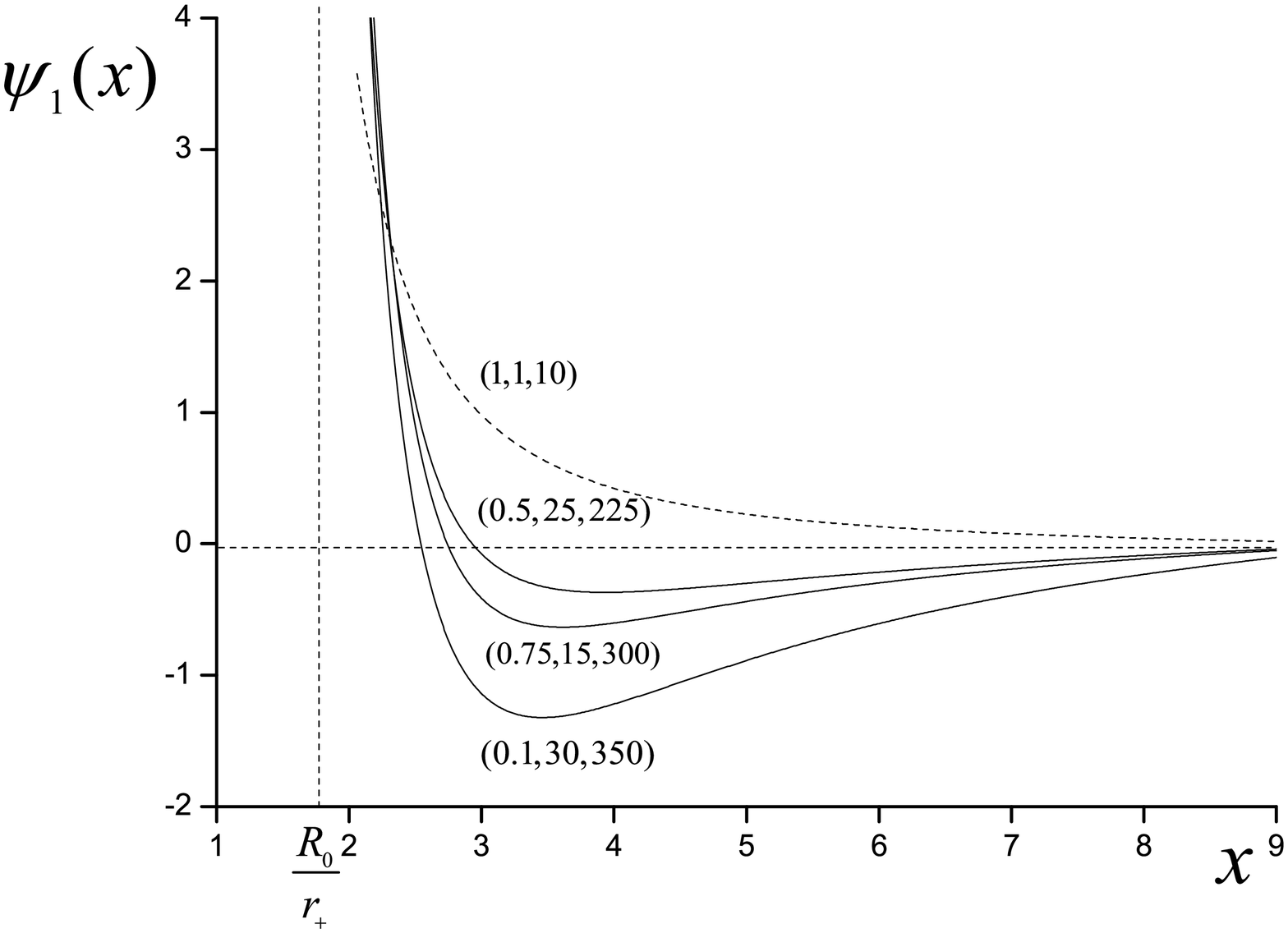}
	\caption{Typical profiles of the dipolar axionic radial function ${\psi_1(x)}$ in the near zone for the case $Q\neq0$. The set of guiding parameters $(a,\omega,\Omega)$  is indicated in parentheses near the corresponding curve. For some sets of the guiding parameters the curves become non-monotonic, and minima appear in the near zone.}
\end{figure}

\subsection{Profile of the axionically induced dipole type electric field}

The equation for the radial function of the dipole type electric field (\ref{key1}) happens to be similar to the (\ref{GR2key}),
$$
F_1^{\prime \prime} + \frac{2}{x} F_1^{\prime}  - \frac{2 F_1}{(x-a)(x-1)} \left[1 + \omega \ \frac{{\cal M}^2(x)}{x^2} \right] =
$$
\begin{equation}
= - \frac{K\Omega {\cal M}(x)}{Qx^2(x-a)(x-1)} \,.
\label{GR28}
\end{equation}
In particular, the part of the function $F_1(x)$, which corresponds to the contribution of axionically induced field, has the same asymptotic behavior as the function $\psi_1$, namely, $F_1 \propto x^{-3}$. As for the behavior of $F_1$ in the near zone ($r \to R_0$), the difference can be explained by the structure of the second term in the left-hand side of  (\ref{GR28}). The profiles of the functions $F_1(x)$  depending on the parameters $a$, $\omega$, $\Omega$ are presented in Fig. 4.

\begin{figure}
	\includegraphics[width=90mm,height=70mm]{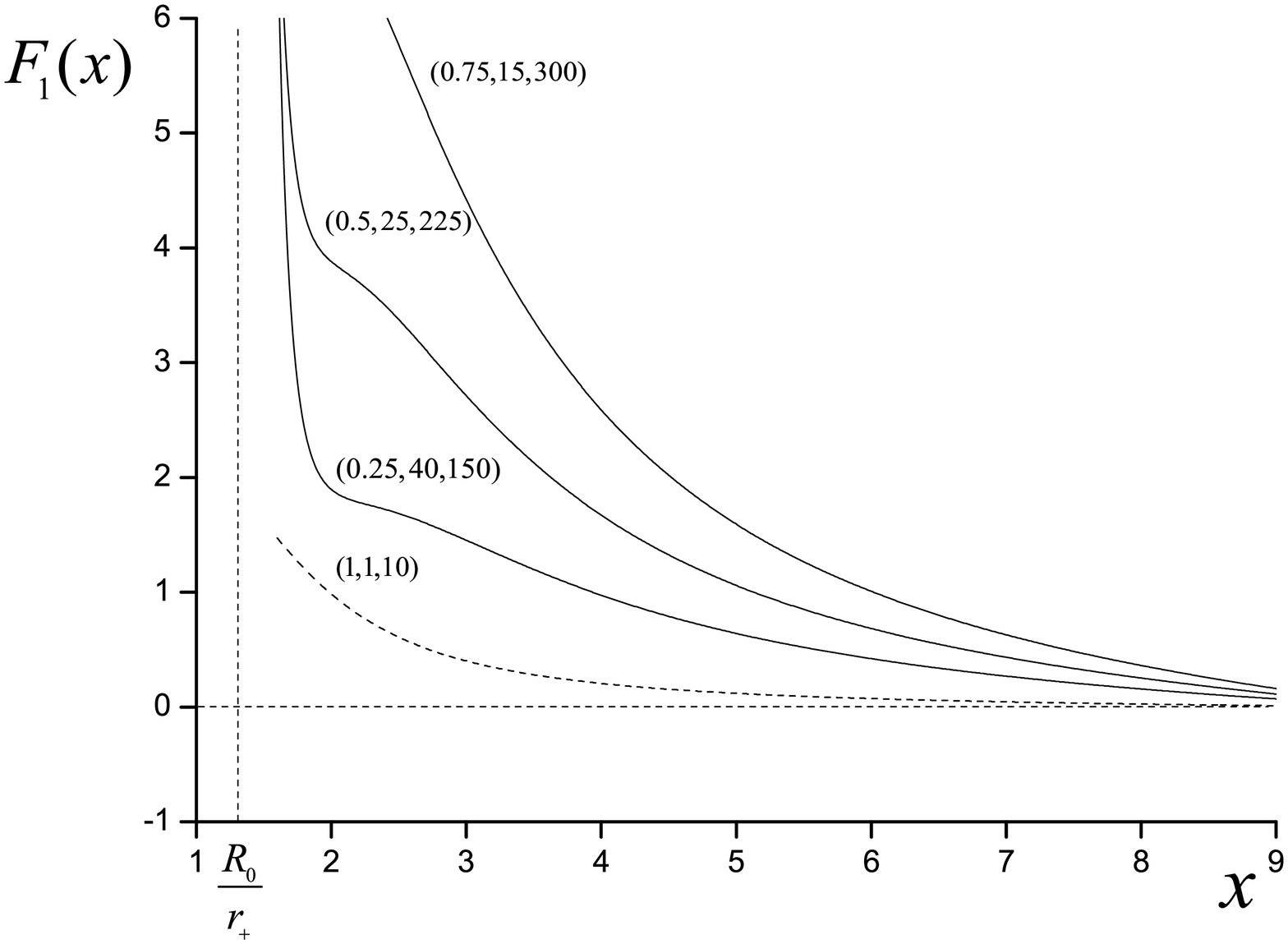}
	\caption{Typical profiles of the axionically induced dipolar electric radial function ${F_1(x)}$ in the near zone for four sets of the guiding parameters, $(a,\omega,\Omega)$. The illustration shows that for some sets of the guiding parameters the curves become non-monotonic and contain the pair of inflection points.}
\end{figure}

\subsection{Restructuring of the monopole type radial functions}

When the functions $G_1$ and $F_1$ are found, we can calculate their contributions to the monopole type radial functions $\psi_0$ and $F_0$. In fact, we now deal with an effective electric charge ${\cal Q}(x)$, which depends on the radial variable and can be represented by the formula
\begin{equation}
{\cal Q}(x) = Q + \frac23 G_1(x) \psi_1(x) \,.
\label{Q3a}
\end{equation}
Similarly, the formula
\begin{equation}
{\cal K}(x) = K - \frac{2}{3\Psi^2_0} G_1(x) F_1(x)
\label{Q3b}
\end{equation}
gives the effective axionic charge ${\cal K}(x)$. Asymptotically, ${\cal Q}(x) \to Q$ and ${\cal K}(x) \to K$ vary rapidly.

\section{Remarks on high order moments}

For the illustration of the idea that the magnetic energy can be redistributed between the first, second, etc., moments under the influence of the axion field, let us consider the submodel for which only two axionic radial functions, $\psi_0$ and $\psi_1$, are nonvanishing; this means that only the monopole and dipole contributions from the axion field are significant. Then the  set of equations for the functions $F_n(r)$ is of the following form:
$$
\left[r^2 F^{\prime}_n(r) \right]^{\prime} - \frac{n(n+1)}{N} F_n =  -  n(n+1) G_n \psi^{\prime}_0  -
$$
$$
  - \left[  \frac{(n+1)^2(n+2)}{(2n+3)} G_{n+1}  + \frac{(n-1)n^2}{(2n-1)} G_{n-1} \right]\psi^{\prime}_1 -
$$
\begin{equation}
 - \left[\frac{(n+1)(n+2)}{(2n+3)} G^{\prime}_{n+1}  - \frac{(n-1)n}{(2n-1)} G^{\prime}_{n-1} \right]\psi_1 \,.
\label{angle311p}
\end{equation}
Similarly, we obtain the equations for the functions $G_m$,
$$
\left(N G_m^{\prime} \right)^{\prime} - \frac{m(m+1)}{r^2} G_m = \psi_0^{\prime}F_m +
$$
\begin{equation}
+\psi_1 \left(\frac{F_{m+1}^{\prime}}{2m{+}3} - \frac{F_{m-1}^{\prime}}{2m{-}1} \right)
+ \sum_{n=0}^{\infty}  h^{(3)}_{m1n} \psi_1^{\prime}F_n \,.
\label{angle32mm}
\end{equation}
In particular, for the quadruple radial functions the equations can be specified as follows:
$$
 \left(r^2 F^{\prime}_2 \right)^{\prime} - \frac{6}{N} F_2 =   -  6 G_2 \psi^{\prime}_0 -
$$
\begin{equation}
 - \left(\frac{36}{7} G_{3}  {+} \frac{4}{3} G_{1} \right)\psi^{\prime}_1 -
  \left(\frac{12}{7} G^{\prime}_{3}  {-} \frac{2}{3} G^{\prime}_{1} \right)\psi_1 \,,
\label{angle311}
\end{equation}
$$
\left(N G_2^{\prime} \right)^{\prime} - \frac{6}{r^2} G_2 = \psi_0^{\prime}F_2 +
$$
\begin{equation}
+\frac47 F_3 \psi_1^{\prime}  + \frac13 F_1 \psi_1^{\prime}   +
 \frac{1}{7} \psi_1  F_{3}^{\prime} - \frac{1}{3}\psi_1  F_{1}^{\prime}
\,.
\label{angle3245}
\end{equation}
Clearly, the master equations are mixed: the dipole function $G_1$ contributes to the equation for the monopole function $F_0$ and the quadruple function $F_2$; the quadruple function $G_2$ contribute to the equation for the dipole function $F_1$, etc. This means  that, the total energy of the magnetic field is redistributed between the multipoles. Clearly, this redistribution is provided by the axion field; to be more precise, the "driving force" of this redistribution is the distortion of the axionic configuration and the appearance of the non-vanishing dipolar radial function $\psi_1 \neq 0$.

\section{Discussion}

The analysis of the model described above allows us to consider three scenarios, following to which the static massive object with pure dipolar magnetic field can be converted into the axonic star with a multipolar magnetic structure, and a homogeneous distribution of the dark matter axions can be converted into a distorted halo, the density of which depends both on the radial variable $r$ and angular variable $\theta$. We assume that this process is very slow, and the evolutionary model consists of a series of states, each being described by static models.

{\it The first scenario} is based on the assumption that initially in the domain near the star with the dipolar magnetic field, there exists a spatially homogeneous background distribution of the axionic dark matter. Until the gradient of the dark matter density is vanishing, the axion-photon coupling is inactive, and the dipolar magnetic field remains undistorted. Then, due to the gravitational attraction the axions drift toward the center of the star, and the axion field becomes the function of the radial variable, $\phi_0(r)$, and its gradient is no longer vanishing ($\nabla_k \phi_0 \neq 0$). In this new situation the axion-photon coupling switches on, and the axionically induced dipolar electric field appears. As a consequence, the interaction between the initial magnetic and induced electric fields produces an additional axion field; it now contains not only the monopole type contribution depending on the $r$ only, but the contribution depending on the meridional angle $\theta$ also. This fact follows directly from the structure of the master equations studied in our work: these equations do not admit solutions in the form of a pure monopole axion field, if the magnetic field is dipolar. Then the interaction between dipolar magnetic and axion fields produces the quadruple component of the electric field, the interaction between dipolar electric and axion fields produces the quadruple component of the magnetic field, and similarly, the quadruple component of the axion field happens to be produced by the coupling of the magnetic and electric fields. This process forms the multipolar structure of the magnetic, electric, and axion fields around the star. Clearly, every additional  $n{+}1$-type moment of the induced field is less by amplitude than the corresponding $n$-type moment, since every step of this sequential procedure adds the multiplier $\frac{1}{\Psi_0}=g_{A\gamma\gamma}$, associated with the constant of the axion-photon coupling
$g_{A \gamma \gamma} < 1.47 \cdot 10^{-10} {\rm GeV}^{-1}$ \cite{CAST}.
Also, mention should be made that the asymptotic profile of the $n{+}1$-type moment differs from the $n$-type one by the multiplier $\frac{1}{r}$: i.e., the $n{+}1$-type magnetic, electric and axionic structures are more compact than the $n$-type ones.

{\it The second scenario} assumes that initially the star with dipolar magnetic field has an electric charge $Q$  and the corresponding radial electric field. Such a configuration inevitably creates axions, and their distribution is inhomogeneous and inherits the features of the electric and magnetic fields profiles. A further part of the second scenario is similar to the corresponding part of the first scenario.

{\it The third scenario} can be realized, if the magnetic star and its axionic environment is situated not far from the black hole (e.g., not far from the super-massive black hole, which forms the active center of a galaxy).
In this configuration the axionic dark matter is already inhomogeneous due to the attraction to the black hole, and the axion-photon coupling is already active. The difference between this scenario and the two previous is that the new configuration of the magnetic and electric fields is not spherically symmetric, and thus, the redistribution of the magnetic energy takes place in much a more complicated manner.

 The first and second scenarios predict that a magnetic star distorts the halo of the axionic dark matter, surrounding the star. The prediction that the halo around the axionically active magnetic star is no longer spheroidal could be interesting, e.g., for specialists, which analyze the gravitational lensing phenomena.

In the presented work, we made the first step in the analysis of the magnetic field restructuring driven by the interaction of the dipolar magnetic field with the axionic dark matter, which surrounds the star. We analyzed the solutions to the equations of  magneto-electro-axiono-statics in the static gravitational field of the Reissner-Nordstr\"om type; it was shown that the key equations for this model can be reduced to the Heun and Fuchs equations, which are known in mathematical physics \cite{Heun1,Heun2,Ince,Poole}. An important detail of this analysis is that three dimensionless guiding parameters of the model happen to be encoded in the structure of the distorted halo of the axionic dark matter: we mean the parameter $a=\frac{r_{-}}{r_{+}}$, describing the relative depth of the inner and outer horizons of the magnetic star, and the parameters $\omega$ and $\Omega$ introduced by (\ref{GR23}), which describe the effectiveness of the axion-photon coupling at the presence of the dipolar magnetic field.
The most interesting magnetic stars, the magnetar, are the rotating objects, and in the future we plan to consider the model of axionically induced magnetic restructuring based on the model with a gravitational field of the Kerr type.


\acknowledgments{The work was supported by Russian Science Foundation (Project No. 16-12-10401), and, partially, by the Program of Competitive Growth of Kazan Federal University.}

\section{Appendix: On the stability of the extended model}

The discussion concerning the stability of the whole model can be formally divided into two parts. In the first part we assume that perturbations, which depend on time, radial and angular variables, are of the pseudoscalar origin, and they are connected with a small deviation of the axion field $\phi$ from its fixed value $\Phi_*$, i.e., $\phi \to \Phi_* + \psi(t,r,\theta,\varphi)$. In this first step, we assume that the electric and magnetic fields remain unperturbed and static. Below we prove that the model is stable under the influence of perturbations of this type.

The second part of the discussion is based on the assumption that the perturbations are of the electromagnetic origin. This means that the Maxwell equations are considered to be non-static, and the axion field to be influenced by the electromagnetic source $-\frac{1}{4\Psi_0}F^{*}_{mn} F^{mn}$. This second step of discussion is beyond the scope of this paper, and we hope to return to this problem in the next work.

\subsection{Scheme of analysis  of the perturbation dynamics}

Let the axion field be presented as $\phi = \Phi_{*} + \psi$, where  $\psi$ is  small.
In the zero order limit with respect to perturbations, we use the results of the analysis of the Eq. (\ref{ax123}) for the function $\phi = \Phi_{*}$. The equation of the first order in perturbations is of the form,
\begin{equation}
\ddot{\psi} =\frac{N}{r^2}\left[\partial_r(Nr^2 \partial_r \psi) {+} \frac{\partial_{\theta}\left(\sin{\theta} \partial_{\theta} \psi \right)}{\sin{\theta}}  {+} \frac{\partial^2_{\varphi} \psi}{\sin^2{\theta}} \right] \,.
\label{A2}
\end{equation}
First of all, we decompose $\psi$ in the series with respect to the spherical functions $Y_{nm}$,
\begin{equation}
\psi(t,r,\theta,\varphi) = \sum_{m,n} \psi_{nm}(t,r) Y_{nm}(\theta,\varphi) \,.
\label{A3}
\end{equation}
For the spherical mode amplitudes $\psi_{nm}(t,r)$, we obtain the equations,
\begin{equation}
\ddot{\psi}_{nm} = \hat{\cal D}\psi_{nm}  \,,
\label{A4}
\end{equation}
where the radial differential operator $\hat{\cal D}$ is of the form,
\begin{equation}
\hat{\cal D} =  \frac{N}{r^2}\left[\partial_r(Nr^2 \partial_r )  - n(n+1)\right] \,.
\label{A5}
\end{equation}
According to the theory of linear operators, we search for the eigen-functions ${\cal U}_{nm}(r)$ and for the corresponding eigen-values $\lambda_{nm}$, which satisfy the equations
\begin{equation}
\hat{\cal D} {\cal U}_{nm} = - \lambda_{nm} {\cal U}_{nm} \,,
\label{A6}
\end{equation}
and the homogeneous boundary conditions,
\begin{equation}
{\cal U}_{nm}(r_0)=0 \,, \quad {\cal U}_{nm}(\infty) = 0 \,, \quad r_0>r_{+}\,.
\label{A7}
\end{equation}
We are faced with the typical Sturm-Liouville problem,
\begin{equation}
\frac{d}{dr}\left[p(r)\frac{d {\cal U}_{nm}}{dr} \right] - q(r) {\cal U}_{nm} + \lambda_{nm} \rho(r) {\cal U}_{nm} = 0,
\label{A8}
\end{equation}
with positive functions $p(r)$, $q(r)$, $\rho(r)$,
$$
p(r) = r^2 N(r) >0 \,, \quad \rho(r) = \frac{r^2}{N(r)} >0 \,,
$$
\begin{equation}
q(r)= n(n+1) >0 .
\label{A9}
\end{equation}
According to the classical results (see, e.g., \cite{Ince}, Chap. X) all the eigen-values $\lambda_{nm}$ are positive for this case, and can be rewritten as $\lambda_{nm}=\omega^2_{nm}$.

The spectrum of the eigen-values can be discrete or continuous; in the discrete case, we add the index $(j)$ to these quantities, and decompose the mode amplitudes ${\psi}_{nm}(t,r)$ with respect to the eigen-functions ${\cal U}_{nm(j)}(r)$ standardly as
\begin{equation}
{\psi}_{nm}(t,r) = \sum_{j} T_{jnm}(t) \ {\cal U}_{nm(j)}(r) \,,
\label{A10}
\end{equation}
and obtain the equations  for the functions $T_{jnm}(t)$,
\begin{equation}
\ddot{T}_{jnm}  + \lambda_{nm(j)} T_{jnm} = 0 \,.
\label{A11}
\end{equation}
If the spectrum is continuous, one has to use the integral with respect to $\lambda$ instead of the sum,
\begin{equation}
{\psi}_{nm}(t,r) = \int d\lambda T_{nm}(t,\lambda) {\cal U}_{nm}(r,\lambda) \,,
\label{A12}
\end{equation}
but the further idea is exactly the same.
Clearly, when the eigen-values $\lambda_{nm}$ are positive, i.e., $\lambda_{nm} = \omega^2_{nm}$ the solutions to the Eq. (\ref{A11}) are the following restricted harmonic functions:
\begin{equation}
T_{nm} = C^{(1)}_{nm} \cos{\omega_{nm}t} + C^{(2)}_{nm} \sin{\omega_{nm}t} \,,
\label{A13}
\end{equation}
and the perturbations have no instable modes. As usual, the coefficients $C^{(1)}_{nm}$ and $C^{(2)}_{nm}$ are predetermined by the initial data for the perturbations. In other words, the key question of the discussion about the stability of the model under the perturbations of the first type is solved: the model is stable with respect to fluctuations of the pseudoscalar field. We add to this discussion two limiting cases as illustrations of the key statement.

\subsection{On the behavior of perturbations of the axion field in the far zone}

In the asymptotic regime, when $r \to \infty$ and thus, $N \to 1$, the differential operator can be simplified as
\begin{equation}
\hat{\cal D} \to  \partial^2_r + \frac{2}{r} \partial_r \,.
\label{A14}
\end{equation}
The leading order terms in the eigen-functions are known to be of the form,
\begin{equation}
{\cal U}_{nm} \propto \frac{1}{r} e^{ikr} \,,
\label{A15}
\end{equation}
the spectrum  is continuous, and the eigen-values,
\begin{equation}
\lambda_{nm} = \omega^2_{nm} = k^2 >0
\label{A16}
\end{equation}
are positive.  There are no instable modes, thus, far from the center of the object the $nm$ modes of perturbations do not grow, and oscillate with the real frequency $\omega = k$.

\subsection{On the behavior of perturbations of the axion field in the near zone}

When we consider the solutions near the outer horizon $r=r_{+}$, we keep in mind that $N(r_{+})=0$ and obtain that the spatial differential operator (\ref{A5}) degenerates, $\hat{\cal D} \to 0$. This means that we have to put the eigen-values to zero, and we see that near the horizon the perturbations are frozen.

\end{document}